# AI- Enhanced Stethoscope in Remote Diagnostics for Cardiopulmonary Diseases

Hania Ghouse [a)], Juveria Tanveen[b)], Abdul Muqtadir Ahmed, and Uma N. Dulhare

*Department of Computer Science and Artificial Intelligence, Muffakham Jah College of Engineering and Technology, Hyderabad, India.*

[a)] haniaghouse704@gmail.com
[b)] tanveen.juveria08@gmail.com

**Abstract.** The increase in cardiac and pulmonary diseases presents an alarming and pervasive health challenge on a global scale responsible for unexpected and premature mortalities. In spite of how serious these conditions are, existing methods of detection and treatment encounter challenges, particularly in achieving timely diagnosis for effective medical intervention. Manual screening processes commonly used for primary detection of cardiac and respiratory problems face inherent limitations, increased by a scarcity of skilled medical practitioners in remote or under-resourced areas. To address this, this study introduces an innovative yet efficient model which integrates AI for diagnosing lung and heart conditions concurrently using the auscultation sounds. Unlike the already high-priced digital stethoscope, this model has been particularly designed to deploy on low-cost embedded devices and thus ensure applicability in under-developed regions that actually face an issue of accessing medical care.This proposed model incorporates MFCC feature extraction and engineering techniques to ensure that the signal is well analyzed for accurate diagnostics through the hybrid model combining Gated Recurrent Unit with CNN in processing audio signals recorded from the low-cost stethoscope. Beyond its diagnostic capabilities, the model generates digital audio records that facilitate in classifying six pulmonary and five cardiovascular diseases. Hence, the integration of a cost effective stethoscope with an efficient AI empowered model deployed on a web app providing real-time analysis, represents a transformative step towards standardized healthcare.

## INTRODUCTION

Non-communicable diseases like cardiovascular and pulmonary diseases are the primary causes of deaths across the world annually, as reported by the World Health Organization [1]. Detection at early stages proves to be important in assuring better results for interventions. In 1816, a French physician, Dr. Laënnec invented a stethoscope, an icon which has remained prominent for primary health check-ups [2, 3]. Although the use of the stethoscope in clinical decision-making has been in practice, the ability of the stethoscope in diagnosing various pulmonary entities still seems ambiguous and inaccurate [2]. It might be due to the reason that sound signals usually become complicated and lead to misinterpretation, thus increasing the risk of misdiagnosis. The problem of identification and interpretation of auscultation sounds has long been admitted in clinical practice [4]. Recent breakthroughs in deep learning have made the possibility of automation in abnormal sound detection, thus easing early diagnosis and screening efforts. Most models are developed for either heart or lung sounds; thus, a combined approach and affordable AI solutions are pivotal for reaching areas with limited health access. The proposed new hybrid model addresses all these challenges and diagnose heart conditions as well as lung conditions using CNN along with GRU to detect the extracted Mel frequency cepstral coefficients from the auscultation sounds. This model is comparatively better than the previous models that took fewer classes, such as Yaseen et al. [5], and Alqudah et al. [6] for heart sounds, and Pham et al. [7], and Fraiwan et al. [8] for lung sounds. Moreover, this system relieves the overfitting problem through ensemble learning similar to the previous work done by Rocha et al. [9]. Thus, a low cost AI-powered stethoscope would considerably reduce human error when integrated with healthcare systems and increase the accuracy of diagnoses, bringing it as one of the salutary instruments in resource-limited environments.

# RELATED WORK

## 1. Cardiac Sound Analysis

The classification and recognition of heart conditions have been studied thoroughly by artificial intelligence experts over time. Early research focused on traditional machine learning techniques, including Karar et al.'s rule based classification tree [10], Sun et al.'s SVM-based ventricular septal defect diagnosis method [11], and Cheema and Singh's naïve Bayes based electrocardiogram grating method [12]. On average, 94% of these machine learning techniques were successful in detecting abnormal heart sounds with a good level of accuracy. Neural network techniques were later proposed, for instance the work done by Sharma et al. [13]. The accuracy rate was about 80% on average. CNN-based techniques have been developed recently. Deperlioglu [14], for example, proposed an eightlayer CNN and obtained 97.90% accuracy. Yaseen et al. [5] used the discrete wavelet transform (DWT) and the mel frequency cepstral coefficient (MFCC) to obtain features from cardiac sound data, as well as a hybrid SVM-DNN model. Over 87% accuracy was attained by their model. Last but not least, Alqudah et al. [6] created a novel approach that produced a 93.70% accuracy rate utilizing the CNN classification algorithm and bispectrumhigher order spectral analysis.

## 2. Pulmonary Sound Analysis

Rocha et al. [15] was among the pioneers in the field of automated lung sound diagnosis. After extracting sound features (such as wheezes, crackles, or both), they classified the data using machine learning algorithms. There are two obstacles to this, however.The distribution of lung sound data is usually skewed across different groups and is rare. Getting valuable information from mild breath sounds is the second challenge. To address the first concern, current literature by Mikolajczyk et al. [16]; Nguyen et al. [17]; and Bardou et al. [18] have used data augmentation techniques. By addressing the issue of class imbalance, these techniques not only increase the model's training data but also enhance its forecast accuracy and applicability across datasets. Bardou et al. [18], for example, employed a large CNN model to achieve the highest acceptable classification accuracy of nearly 91%. Various feature extraction strategies have also been developed in prior studies to address the second difficulty. Demir et al. [19] For example, employed the short-time Fourier transform method to translate lung sounds into spectrogram images. Continue wavelet transform and empirical mode decomposition were utilized by Shuvo et al. [20].

# EXISTING SYSTEM

Digital stethoscopes are invaluable tools for physicians, enhancing diagnostic capabilities through advanced features like sound amplification, noise reduction, and connectivity options. However, the high costs significantly limit their accessibility, particularly at the individual level in developing nations and underserved communities [21]. The cost of these devices often makes them impractical for their wide use in resource-constrained environments. In many under-developed regions healthcare systems face challenges due to shortages of medical professionals and inadequate infrastructure [22] Though this may be very useful and able to promote remote diagnosis and telemedicine upon adopting digital stethoscopes, financial accessibility has proved to be a big barrier in places where quality health care is already at a low level. Recently developed low-cost versions of digital stethoscopes also focus on lung auscultation while giving due importance to heart sound analysis to an equal degree. This smallness of focus creates a limitation of thorough diagnostic capability that these devices could amply fulfill [23].This study seeks to overcome this hurdle by introducing a hybrid AI model designed to simultaneously assess both lung and heart sounds. By integrating advanced deep learning techniques into an affordable stethoscope, the aim is to enhance diagnostic accuracy while ensuring accessibility for healthcare providers in low-resource settings. This innovative approach could ultimately bridge the gap in care for patients in remote areas, allowing for more effective telemedicine practices and improving overall health outcomes.

# PROPOSED SYSTEM

Although the latest developments in artificial intelligence have greatly improved the diagnosis of heart and lung diseases through auscultation sounds, there are still many challenges faced by the current models. Traditional techniques in machine learning have already reached a satisfactory accuracy level for detecting cardiac and pulmonary conditions. However, the significant challenges like detecting and classifying the heart and lung diseases in more categories, managing and balancing sound data from heart and lung simultaneously, integration of hybrid AI models into embedded devices to ensure practicesin medical diagnostics have open up further research issues. To overcome the above limitations, the proposed light-weight hybrid model combines the capabilities of Convolutional Neural Network and Gated Recurrent Unit (CNN+GRU), as illustrated in Fig1.

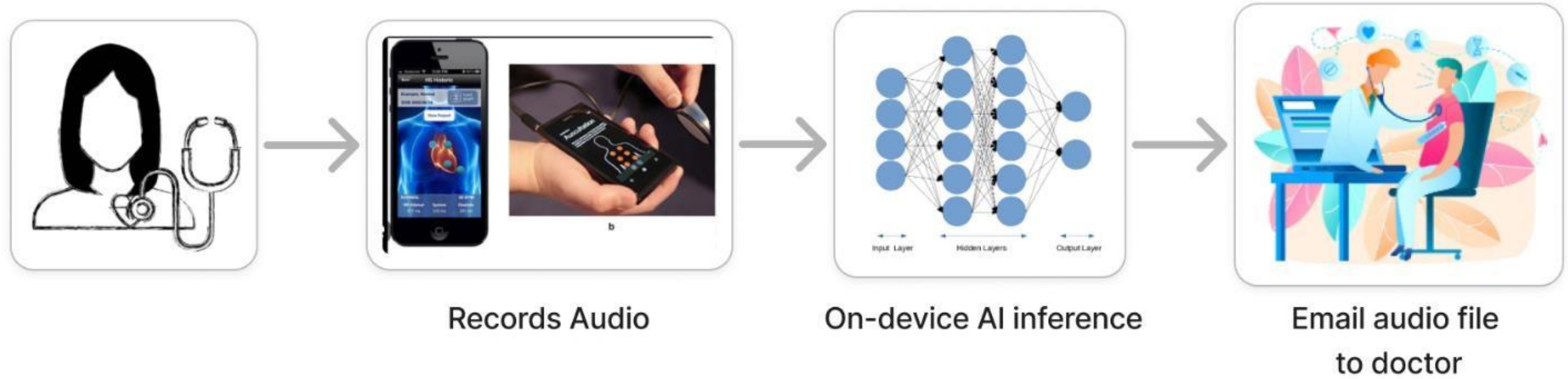

**FIGURE 1**. Application Workflow

## 1. Proposed Hardware Setup

The setup includes a traditional stethoscope to facilitate the auscultatory sounds when placed in contact with a patient's skin near the heart or the lungs. A microphone is connected to the stethoscope that captures the vibrations of the sound waves produced. As shown in Fig 2, the connecting wire secured with a 3.5mm Jack on the other end can be connected to any device. This helps in transferring the auscultatory sounds to a device which hosts the web app for remote diagnosis.

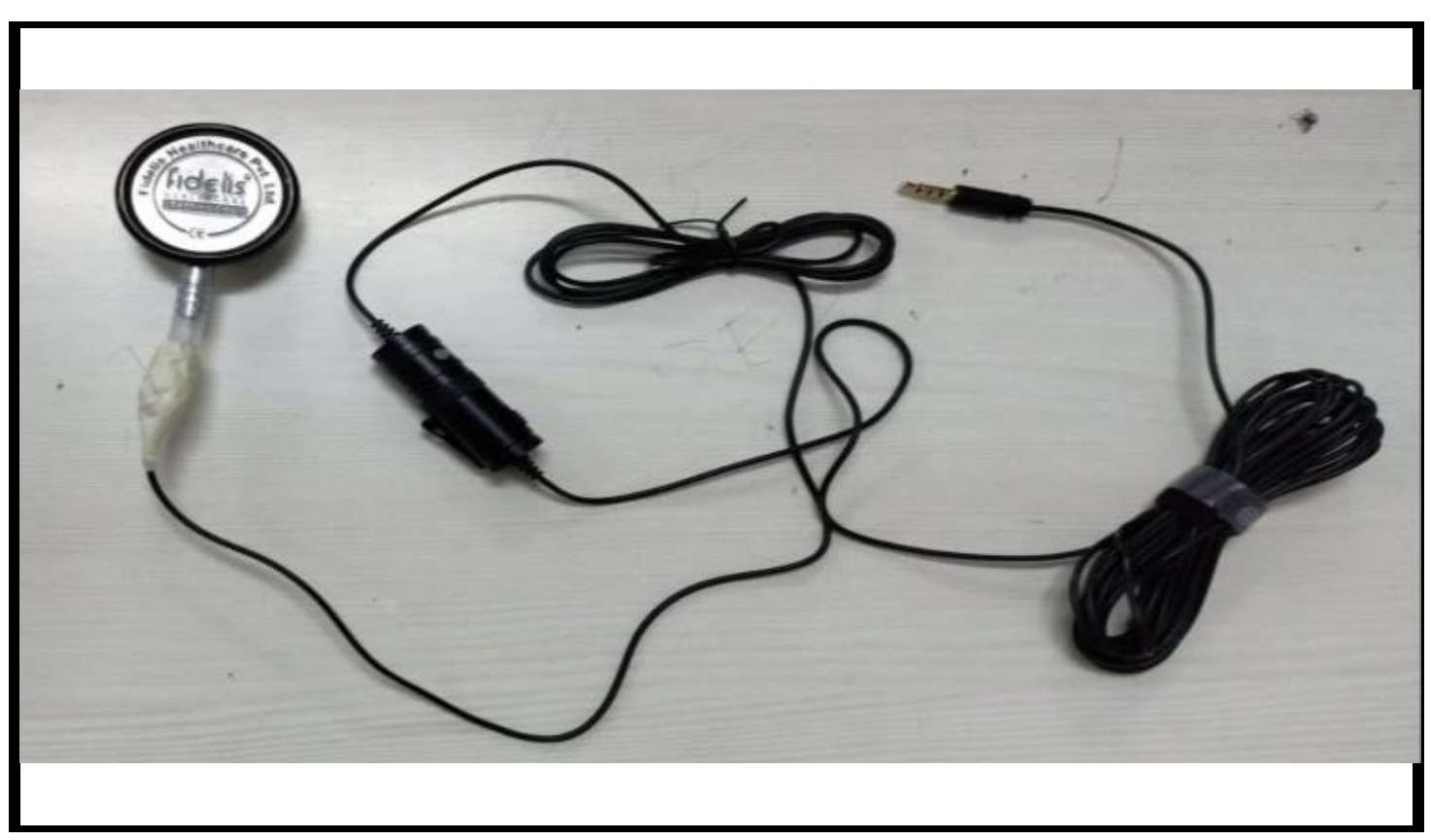

**FIGURE 2**. Proposed Hardware Setup

## 2. Dataset Descriptio

The open-source datasets for cardiac and respiratory auscultation sounds are gathered, as shown in Table 1. The International Conference on Biomedical Health Informatics (ICBHI) 2017 is where lung sound data repository was obtained. The dataset contained the following classifications: pneumonia (P), bronchiectasis (BA), bronchiolitis (BO), upper respiratory tract infection (URTI), chronic obstructive pulmonary disease (COPD), and healthy (H). Using Yaseen et al. [10] heart sound dataset, heart sound diagnosis was carried out. The 1000 sound recordings are distributed uniformly across five primary categories: aortic stenosis (AS), mitral stenosis (MS), mitral regurgitation (MR), normal (N), and mitral valve prolapse (MVP).

**TABLE 1.** Dataset Description before Augmentation

| Types | Class | Original |
|---|---|---|
| **Heart Sounds** | Aortic Stenosis (AS) | 200 |
| | Mitral Stenosis (MS) | 200 |
| | Mitral Regurgitation (MR) | 200 |
| | Normal (N) | 200 |
| | Mitral Valve Prolapse (MVP) | 200 |
| **Lung Sounds** | Chronic Obstructive Pulmonary Disease (COPD) | 793 |
| | Pneumonia (P) | 37 |
| | Bronchiectasis (BA) | 16 |
| | Bronchiolitis (BO) | 13 |
| | Healthy (H). | 35 |
| | Upper Respiratory Tract Infection(URTI) | 23 |

## METHODOLOGY

Despite prior machine learning models demonstrated good outputs,further improvements are yet to be explored. First off, no model can handle heart and lung sound data at the same time in any of these. Secondly, the size of these models limit their effectiveness for embedded device deployment. Finally, adding a new classifier can improve these models' classification performance even further. The goal is to develop an optimized integrated model that combines the advantages of meta-learning with GRU and CNN. The trials showed that the hybrid model can be implemented on a low-cost device to accurately diagnose five different heart disease types and six different lung disease types. The overall working of the system is divided into steps as shown in Fig 3 and explained in subsequent sections.

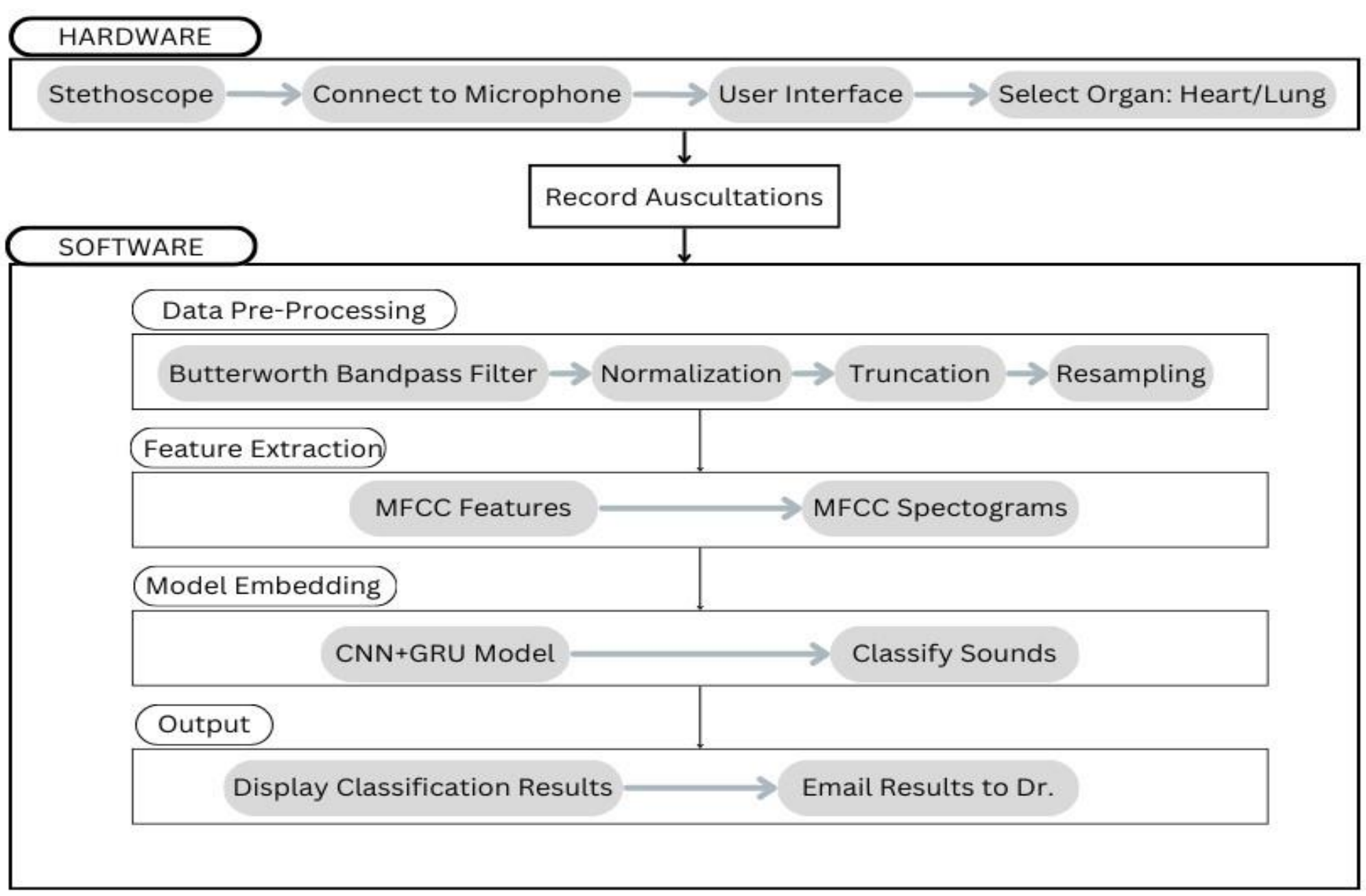

**FIGURE 3**. Overall System Implementation

## 1. Data Pre-Processing

As has been noted in earlier research, the typical frequency range for auscultation signals is 25–400 Hz. Data files, or signal sequences, from the pulmonary and cardiac acoustics datasets were initially processed, as illustrated in Fig 3, with a 2nd-order Butterworth band pass filter, which had upper and lower cut-off frequencies of 25 and 400 Hz, respectively. The sound signals samples is then rescaled at 1000 Hz to uphold stability and minimize processing expenses. After that, the length of every sound signal sequence—that is, the first 2500 data points—was lowered to 2.5 seconds. The impact of variation was reduced by normalizing each signal sequence to (−1, 1). To enhance the range and volume of a training data, data augmentation is a commonly used machine learning technique. This process involves loading audio files as waveforms and applying augmentation to the waveform. In the context of the lung sound dataset, ICBHI 2017 suffers from imbalance problems, where some classes have far fewer samples than COPD. Augmentation helps balance this problem of class imbalance by generating artificial data on underrepresented classes. This also helps avoid bias toward one particular class that the model may take. Also, to keep the augmented data relevant, it is important to ensure it retains the same labels as the original samples. Some augmentation methods include adding artificial noise to the audio signals, temporal shifting of the audio signals, and alteration in the pitch and speed of the audio signals.

## 2. Feature Extraction and Mel Spectrogram Generation

In analyzing audio signals, Mel spectrograms and Mel-Frequency Cepstral Coefficients (MFCCs) are essential elements in analyzing auscultation. MFCCs are very effective in case of audio. Feature extraction, which is the second step after data preprocessing as indicated by Fig 3, increases the efficiency of audio classification of respiratory sounds in the presence of a number of disturbing factors, such as background sounds and physiological artifacts. The first stage amplifies the high frequency components to increase the signal-to-noise ratio. Then, audio signal is taken through a Fourier transform that essentially takes it out from the time domain into the frequency domain and can be represented by its component frequencies. This is followed by windowing, where the audio is cut into small segments that overlap each other for frequency analysis over a short duration, as shown in Fig 4. For every segment, the power of the signal is spread over the various frequency components in what is referred to as the magnitude spectrum. This is then followed by adopting a log of the present frequency axis to what is called as the Mel-Spectrogram.

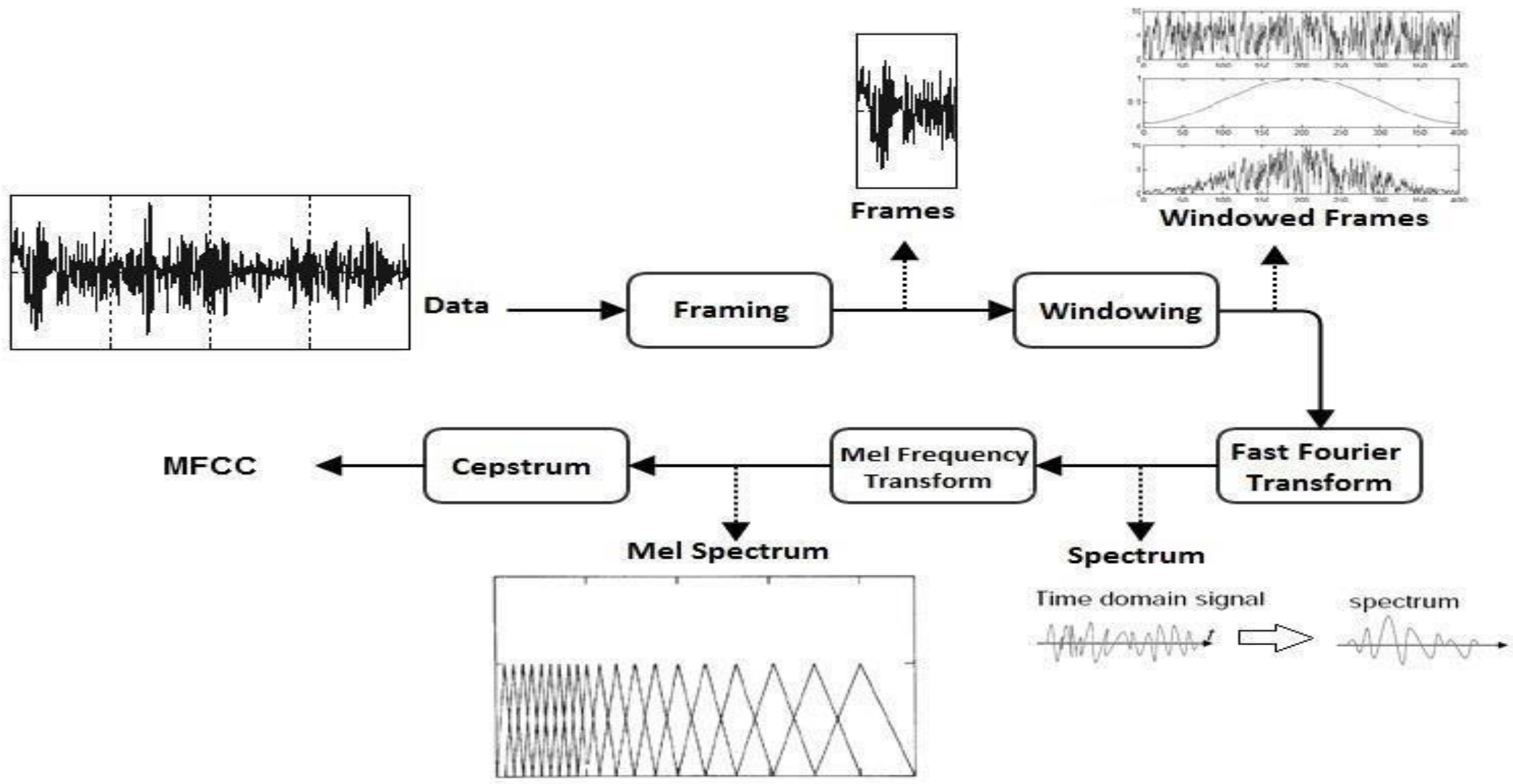

**FIGURE 4.** MFCC Generation

To generate MFCCs, a series of Mel filter banks captures energy across various frequency bins, mimicking human auditory response. Logarithmic compression is applied to the filter bank energies, and finally, these energies are transformed into the cepstral domain using the Discrete Cosine Transform (DCT), producing the MFCCs that encapsulate essential audio features, as shown in Fig 4. A visual comparison of Mel spectrograms before and after noise reduction, highlighting the effectiveness of preprocessing techniques in enhancing audio signal quality for better interpretation and analysis can be seen in Fig 5.

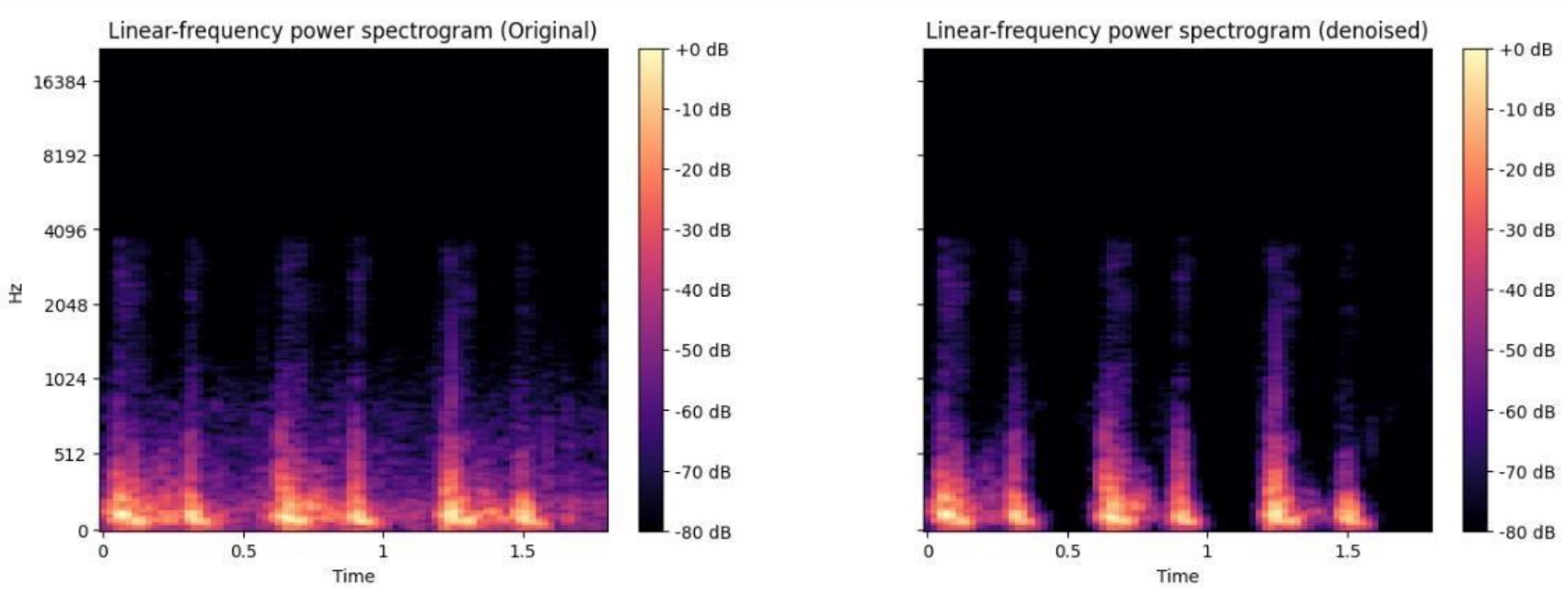

**FIGURE 5**. Mel Spectrogram before and after De-noising

While spectrograms show the evolution of frequencies over time directly, MFCCs provide a compact representation that emphasizes the features most relevant to human auditory perception. Spectrograms are used for visualizing and analyzing sound, whereas MFCCs are typically used as input features for machine learning models in audio processing tasks. The labels are encoded into one-hot format for use in the neural network.

## 3. Model Architecture

The next step after MFCC feature extraction is developing a machine learning model to classify the auscultations accurately, as shown in Fig 3. The model architecture presented herein combines Convolutional Neural Networks (CNNs) and Gated Recurrent Units (GRUs) to effectively process audio data. The input layer accepts MFCC features with a shape of (1, 52). It comprises two convolutional layers: the first with a kernel size of 11, 256 filters, followed by max pooling (2x2), batch normalization, and ReLU activation, and the second with 512 filters and similar configurations. This is followed by five sets of GRU layers, featuring varying units (32, 64, and 128) and employing 'tanh' activation, where the outputs from these sets are combined in an additive layer. The dense layers consist of intermediate sets using Leaky ReLU activation, culminating in a final output layer with 11 units utilizing softmax activation for classification. This architecture, illustrated in Fig 6, leverages the strengths of both CNNs and GRUs to enhance audio feature extraction and sequential learning.

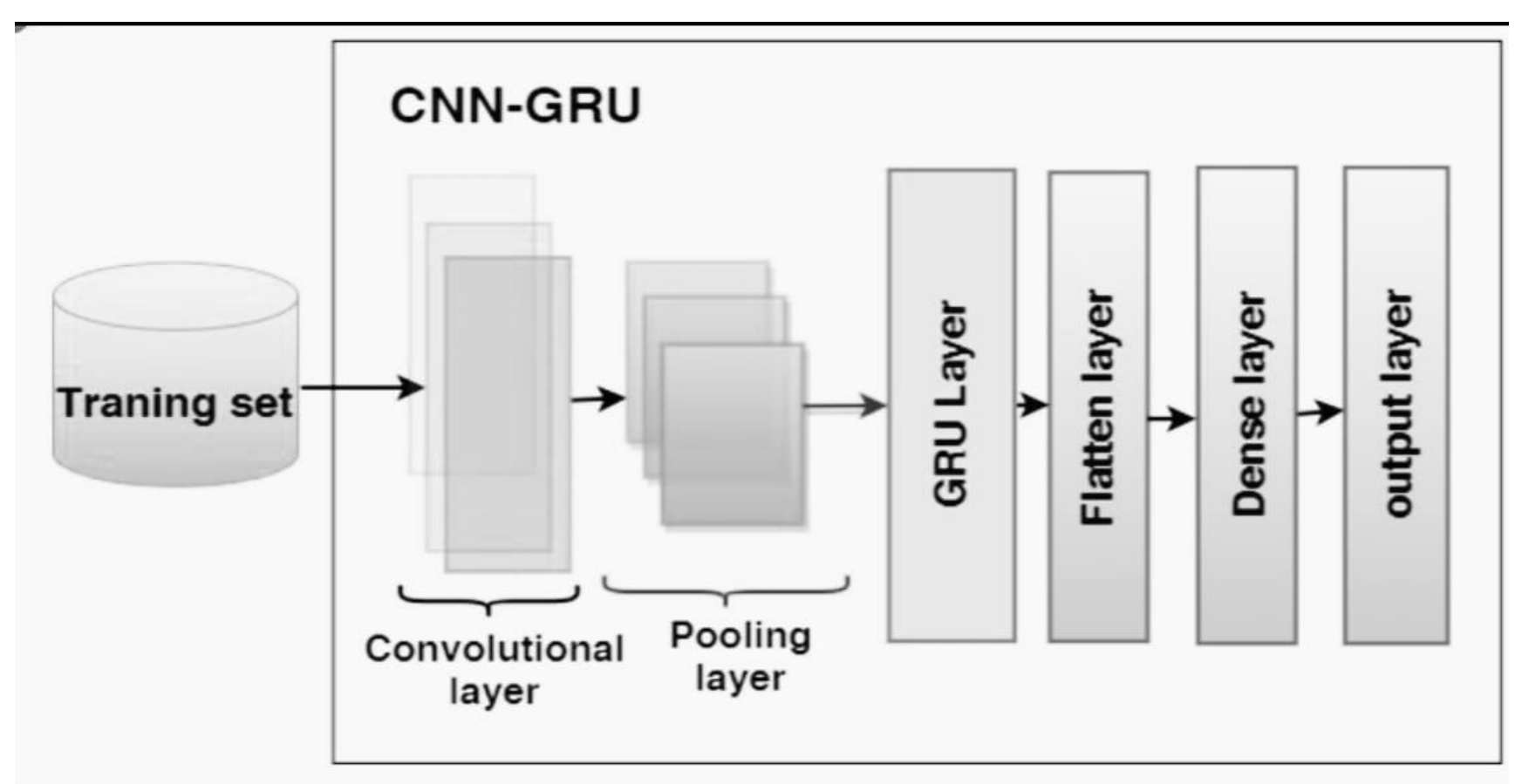

**FIGURE 6.** Model architecture of the proposed hybrid CNN-GRU model

## 4. Training Process

Model compilation involves configuring the learning process with the Adam optimizer, set at a learning rate of 0.0002, used for its efficiency in training deep learning models by combining advantages of AdaGrad and RMSProp. The Categorical Crossentropy loss function is applied for multi-class classification, with model performance assessed using accuracy, precision, F1-score, and recall. To enhance training, callbacks such as Early Stopping and Model Checkpoint are implemented; Early Stopping halts training if validation accuracy does not improve after 300 epochs, reverting to optimal weights to prevent overfitting, while Model Checkpoint saves model weights whenever validation performance improves. The model is trained with a batch size of 8 over 150 epochs, where the batch size indicates the number of samples per training iteration and the epochs reflect the number of complete passes through the training dataset. During this phase, validation data is employed to monitor generalization to new data.

## 5. Testing Process

The dataset is split into three subsets: the Train data, which is used for model training and further split to create the Validation Set, making the final size the remainder after this split; the Validation Set accounts for 17.5% of the original data and is used to assess the model performance and fine-tune hyperparameters; and the Test Set, comprising 7.5% of the data, is reserved for assessing the final model's performance. Statistical metrics such as precision, recall, and F1 score are used particularly as they are important for unbalanced datasets, and they give a better detailed assessment than accuracy alone. Based on the performance metrics, the highest accuracy model is then integrated with our front end user interface which will be used to classify the real time heart and lung sounds acquired from the low cost enhanced stethoscope, as illustrated in Fig 3. The lung and heart sound recordings are segmented into either respiratory or cardiac cycles, preprocessed into MFCC features, and then fed into the model for classification. The lightweight design of the trained model enables efficient deployment on embedded devices, allowing real-time diagnostics in low-resource settings.

# RESULTS AND DISCUSSION

To assess the best model for classifying heart and lung auscultations, the proposed hybrid model is employed alongside the CNN and GRU models. For lung sound classes, the CNN model achieved an average accuracy rate of 74.16%, while the GRU model slightly outperformed it with an accuracy of 75.83%, refer Table 2. The results of heart sound classes, as seen in Table 3, the CNN model attained an average accuracy of 81%, and the GRU model achieved 83%. In contrast, the hybrid model showcased a remarkable average accuracy of 94% for both lung and heart sounds, highlighting the combined power of the CNN and GRU networks.

**TABLE 2.** Comparative metric analysis for Lung Sound classes

| Model | Class | Precision | Recall | F1-Score | *Accuracy* |
|---|---|---|---|---|---|
| **CNN** | Bronchiectasis | 0.85 | 0.70 | 0.77 | 0.74 |
| | Bronchiolitis | 0.65 | 0.35 | 0.45 | 0.68 |
| | COPD | 0.90 | 0.88 | 0.89 | 0.85 |
| | Healthy | 0.70 | 0.95 | 0.81 | 0.87 |
| | Pneumonia | 0.55 | 0.50 | 0.52 | 0.65 |
| | URTI | 0.20 | 0.15 | 0.17 | 0.60 |
| **GRU** | Bronchiectasis | 0.83 | 0.72 | 0.77 | 0.75 |
| | Bronchiolitis | 0.60 | 0.40 | 0.48 | 0.70 |
| | COPD | 0.92 | 0.90 | 0.91 | 0.89 |
| | Healthy | 0.78 | 0.90 | 0.84 | 0.80 |
| | Pneumonia | 0.50 | 0.45 | 0.47 | 0.68 |
| | URTI | 0.25 | 0.20 | 0.22 | 0.61 |
| **CNN+GRU** | Bronchiectasis | 0.90 | 0.85 | 0.87 | 0.90 |
| | Bronchiolitis | 0.95 | 0.92 | 0.93 | 0.94 |
| | COPD | 0.82 | 0.78 | 0.80 | 0.85 |
| | Healthy | 0.85 | 0.80 | 0.82 | 0.89 |
| | Pneumonia | 0.65 | 0.60 | 0.62 | 0.75 |
| | URTI | 0.90 | 0.88 | 0.89 | 0.93 |

**TABLE 3**. Comparative metric analysis for Heart Sound classes

| Model | Class | Precision (%) | Recall (%) | F1-Score (%) | *Accuracy (%)* |
|---|---|---|---|---|---|
| **CNN** | AS | 83.12 | 78.56 | 80.81 | 80.52 |
| | MR | 81.34 | 82.47 | 81.90 | 82.03 |
| | MS | 77.89 | 80.15 | 78.95 | 79.52 |
| | MVP | 84.27 | 75.61 | 79.70 | 80.04 |
| | N | 82.68 | 83.21 | 82.94 | 83.02 |
| **GRU** | AS | 80.56 | 79.34 | 79.95 | 79.58 |
| | MR | 76.89 | 81.45 | 79.11 | 80.23 |
| | MS | 78.23 | 75.78 | 76.77 | 76.51 |
| | MVP | 82.14 | 80.92 | 81.52 | 81.09 |
| | N | 79.68 | 84.12 | 81.84 | 83.08 |
| **CNN +GRU** | AS | 93.45 | 94.12 | 93.78 | 94.01 |
| | MR | 94.11 | 93.78 | 93.94 | 94.02 |
| | MS | 93.89 | 93.56 | 93.72 | 94.07 |
| | MVP | 94.34 | 93.67 | 94.02 | 94.05 |
| | N | 93.78 | 94.23 | 94.08 | 94.09 |

# 1. Performance Analysis

According to the confusion matrices shown in Fig 7, the combined CNN and GRU model demonstrates better performance in classifying respiratory and heart conditions compared to the individual CNN and GRU models.

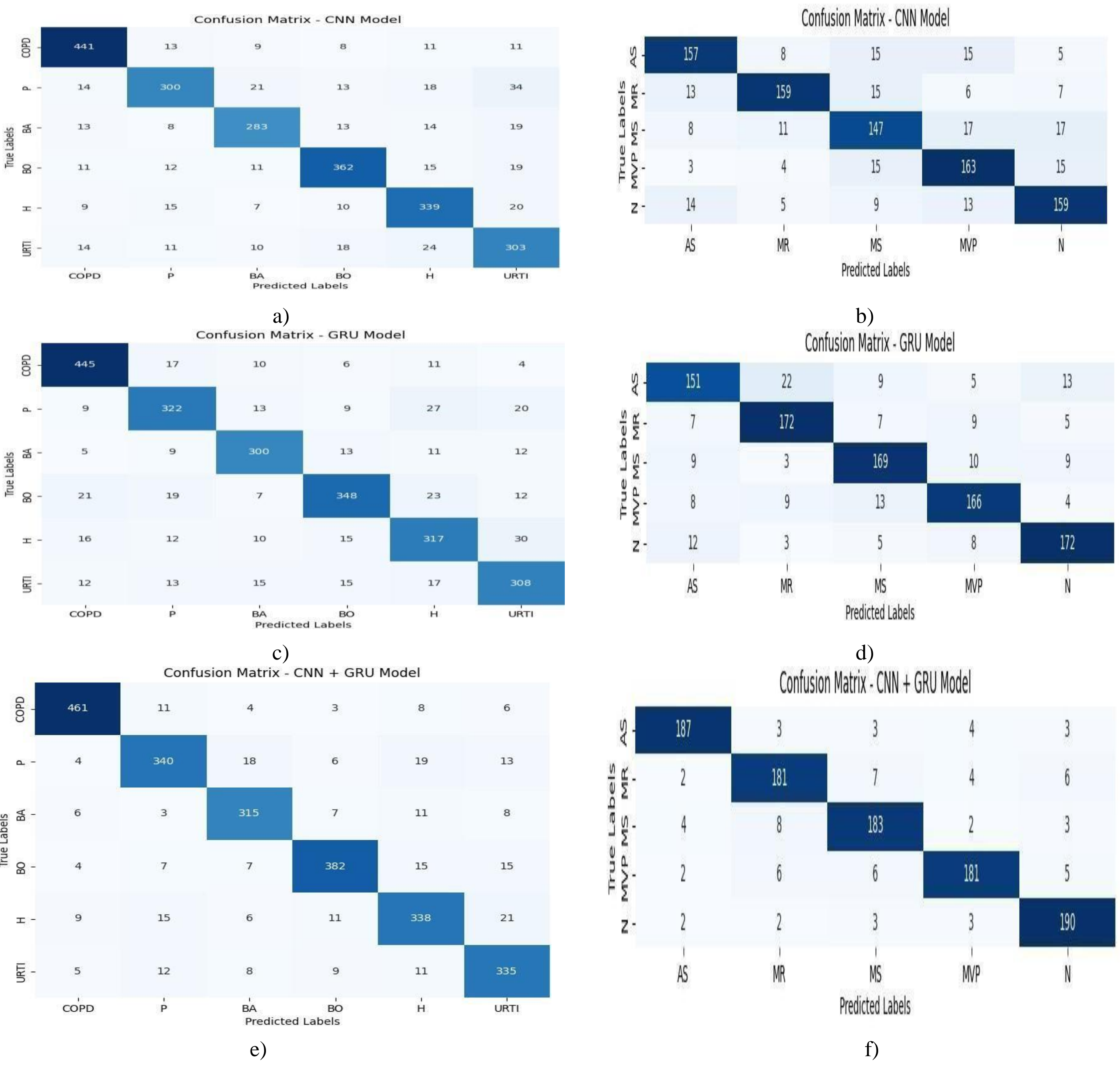

**FIGURE 7.** a) Error matrix for lungs CNN model, b) Error Matrix for heart CNN model, c) Error Matrix for lungs GRU model, d) Error Matrix for heart GRU model, e) Error Matrix for lungs CNN+GRU model, f) Error Matrix for heart CNN+GRU model

This significant improvement in accuracy underscores the effectiveness of hybridizing CNN and GRU networks, which facilitates the extraction of essential features while efficiently utilizing their temporal dependencies. Although the CNN model yields acceptable results when used alone, the performance of the GRU model declines sharply, further emphasizing the advantages of the hybrid approach. Overall, these findings, as shown in Fig 8, indicate that the CNN+GRU model not only improves classification accuracy across various respiratory conditions but also demonstrates a robust capability in analyzing heart auscultations.

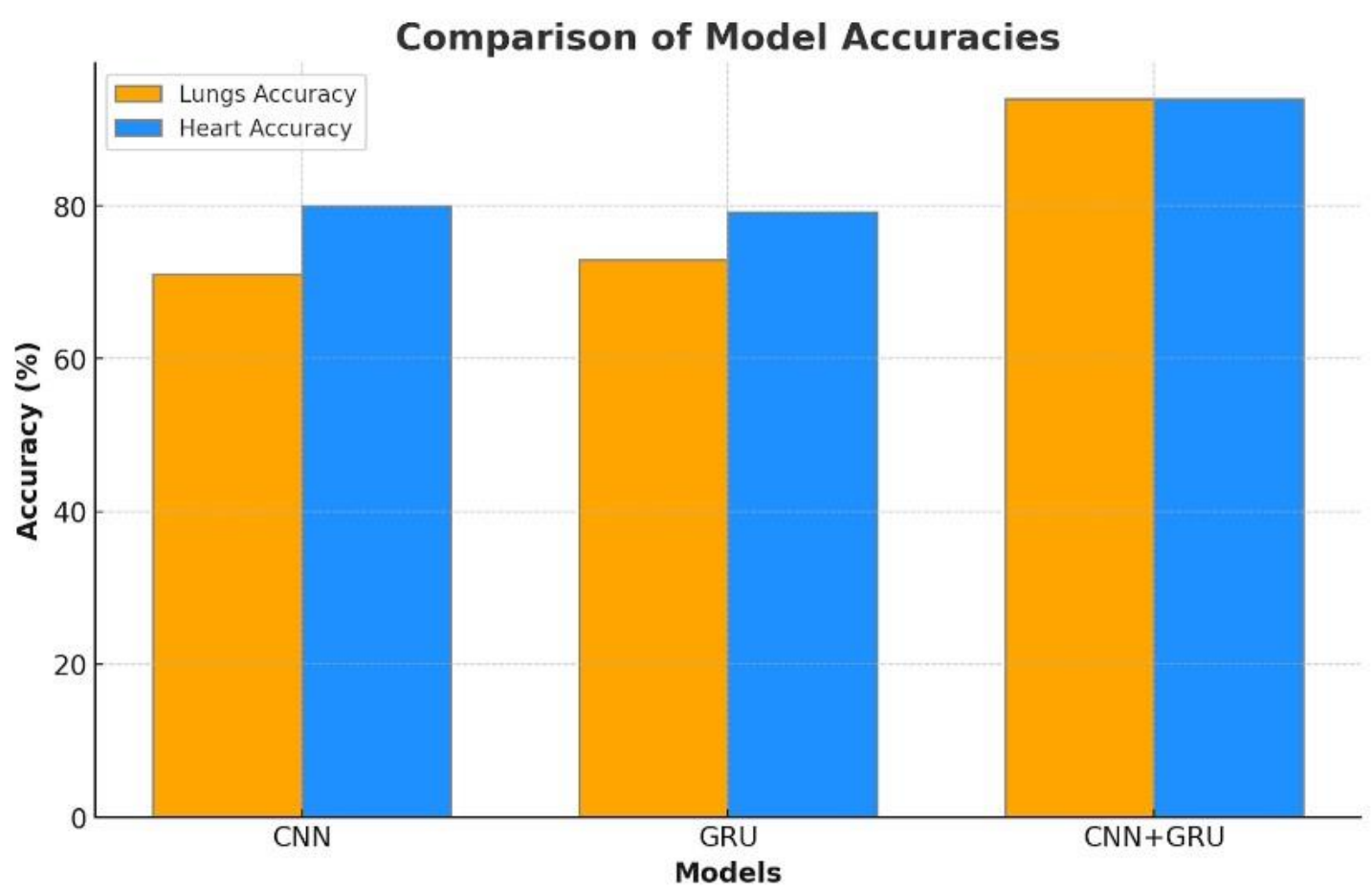

**FIGURE 8**.Comparative Results for the Lung and Heart sounds Model

## 2. System User Interface

The AI-powered stethoscope's user interface (UI) is designed for ease of use, catering to both healthcare professionals and non-experts. It features real-time interaction for starting, stopping, and reviewing recordings, along with a reporting system that captures auscultation sounds and delivers AI-based classifications. Integrated email functionality supports telemedicine by allowing users to send reports to doctors for remote consultations. The application, built using Streamlit, incorporates TensorFlow for real-time data capture and is initially deployed locally before scaling to platforms like Heroku for broader access. Clear instructions, as shown in Fig 9, guide users on recording lung and heart sounds. Users can record auscultation sounds using the device. Users can start and stop recordings and review sounds immediately. Once recorded, and then submitted the UI returns a predicted class for the recorded sounds based on a hybrid model CNN & GRU analysis in the backend. Telemedicine support is provided through this web app which allows the users at the UI to send the generated reports to doctors via email, as illustrated in Fig 3, supporting remote consultations and telemedicine workflows.

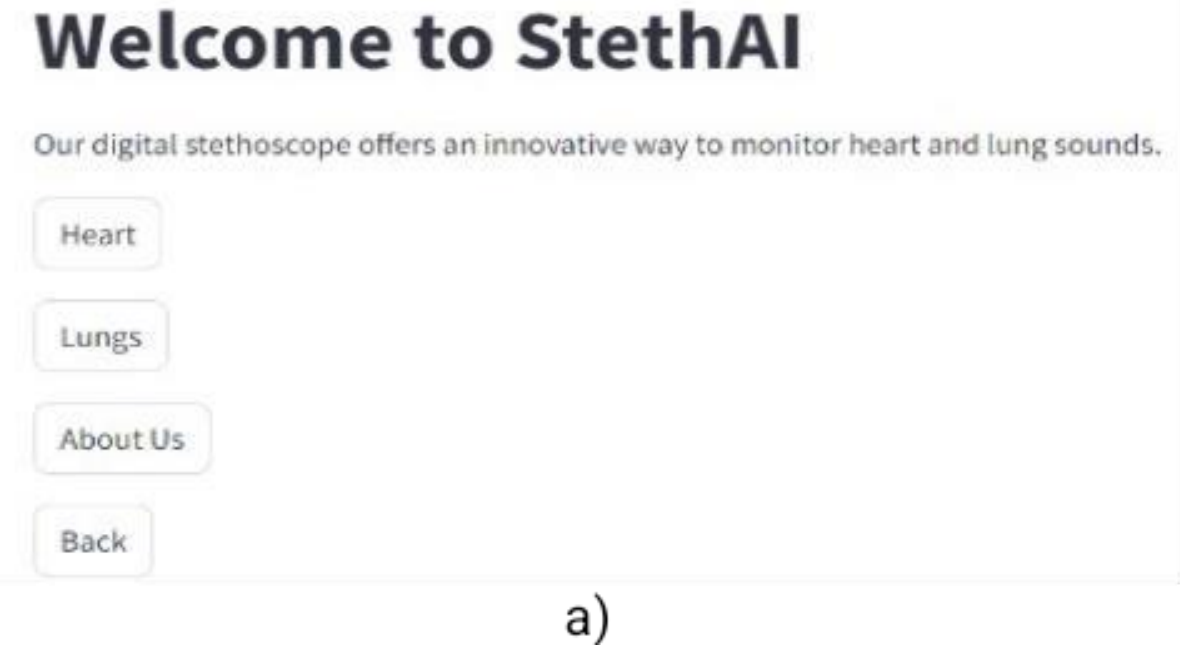

a)

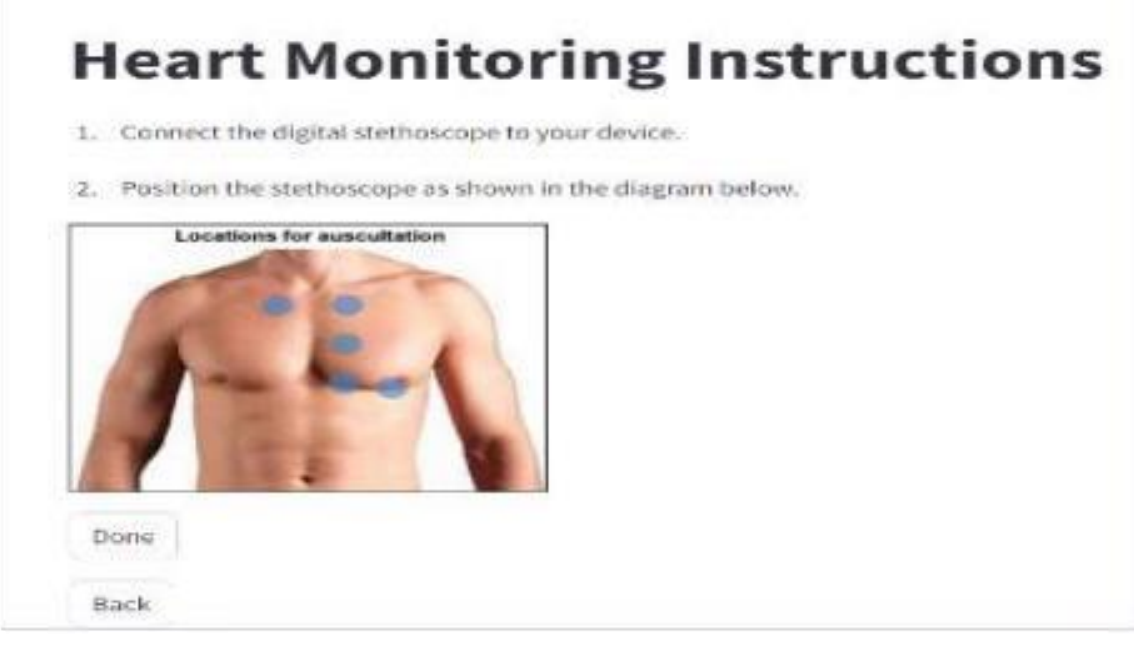

b)

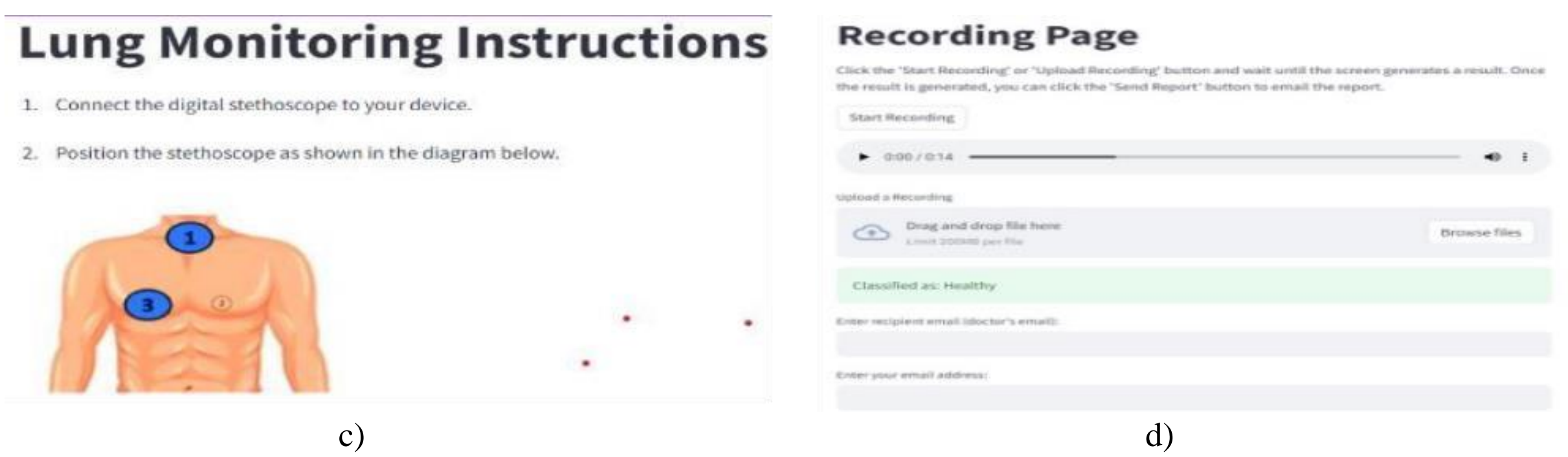

c) d)

**FIGURE 9.** a) Front page of the Web application, b) Redirects to heart instructions page, c) Redirects to lungs instructions page if selected, d) Plays the recording and shows Classification Results

## CONCLUSION & FUTURE SCOPE

In this study, the proposed novel hybrid model is able to diagnose jointly pathologies of the heart and lungs. This model overcomes severe problems resulting from high class numbers and class imbalance. The proposed system handles eleven different classes with an accuracy rate of 94%, outperforming other models dealing with fewer classes. This architecture couples CNN with GRU for a more accurate classification by overfitting, using temporal and frequency-based dependencies. It also is poised for deployment with a web app connected to an affordable stethoscope and the ability to run improved real-time diagnostics on mobile devices. Although the current study shows promising results, large-scale clinical testing, and expansion of data at various hospitals would be in furtherance of the fine-tuning and validating effectiveness for the model.